\documentclass[12pt]{iopart}

\usepackage{iopams}
\usepackage{epsfig,bm,epsf,graphics}
\begin{document}

\title[Phase Transition in Fully Frustrated Lattice]{Flat Energy-Histogram Simulation of the Phase Transition in an Ising Fully Frustrated Lattice}

\author{V. Thanh NGO}

\address{Institute of Physics, P.O. Box 429,   Bo Ho, Hanoi 10000,
Vietnam\\
Division of Fusion and Convergence of Mathematical
Sciences, National Institute for Mathematical Sciences, Daejeon, Republic of
Korea}
\ead{nvthanh@iop.vast.ac.vn}

\author{D. Tien HOANG}

\address{Institute of Physics, P.O. Box 429,   Bo Ho, Hanoi 10000,
Vietnam}
\ead{hdtien@iop.vast.ac.vn}

\author{H. T. DIEP\footnote{Corresponding author}}

\address{Laboratoire de Physique Th\'eorique et Mod\'elisation,
Universit\'e de Cergy-Pontoise,\\ CNRS UMR 8089\\
2, Avenue Adolphe Chauvin, 95302 Cergy-Pontoise Cedex, France}
\ead{diep@u-cergy.fr}

\begin{abstract}
We show in this paper the results on the phase transition of the so-called fully frustrated simple cubic lattice with the Ising spin model. We use here the Monte Carlo method with the flat energy-histogram Wang-Landau technique which is very powerful to detect weak first-order phase transition.  We show that the phase transition is  clearly of first order, providing a definite answer to a question raised 25 years ago.
\end{abstract}

\pacs{05.50.+q 	Lattice theory and statistics , 64.60.Cn Order-disorder transformations ,
75.40.Mg    Numerical simulation studies}
\maketitle

\section{Introduction}

Statistical physics provides powerful methods to study behaviors of systems of interacting particles. In particular, different kinds of transition from one phase to another has been studied with efficiency during the last 40 years by exact methods\cite{Baxter}, renormalization group, high- low-temperature expansions\cite{Zinn}, numerical simulations, ... Experiments have verified most of these theoretical results.
Among the most studied subjects, we mention the effect of the frustration
in spin systems.  The frustration is known to be the origin of spectacular properties such as large ground state (GS)
degeneracy, successive phase transitions, partially disordered phase, reentrance and disorder lines.  Though these aspects have been found in exactly solved models\cite{Diep-Giacomini}, we believe that many of these features remain in complicated frustrated systems where exact solutions are not available.
These general frustrated systems still constitute at present a challenge for theoretical physics\cite{Diep2005}.

Let us recall the definition of a frustrated system.
When a spin cannot fully satisfy energetically all the interactions with its neighbors, it is
"frustrated".   This occurs when the interactions are in competition with
each other, for instance incompatible nearest-neighbor (NN) and next-nearest-neighbor (NNN) interactions,   or when the lattice geometry does not allow a spin to satisfy all interaction bonds simultaneously such as the triangular antiferromagnet.
Except a few two-dimensional frustrated Ising systems where exact methods have been devised to solve with mathematical elegance\cite{Diep-Giacomini,Diep1987,Diep1989,Diep-Debauche91,Diep1991a,Diep1992}, most systems have recourse to numerical simulations and various approximations.
One of the most studied systems is the stacked triangular antiferromagnet (STA) with interaction between NN. This system with Ising\cite{Diep93},  XY and Heisenberg spins\cite{Delamotte2004,Loison} have been intensively studied since 1987\cite{kawamura87,kawamura88,azaria90,Loison94,Boub,Dobry,antonenko,Loison2000}, but only recently that the 20-year controversy comes to an end\cite{tissier00b,tissier00,tissier01,itakura03,Peles,Kanki,Bekhechi,Zelli,Ngo08,Diep2008}.  Note that numerical simulations require now new efficient algorithms to deal with frustrated systems\cite{Ngo08,Diep2008}.

There is another fully frustrated system.  Initially defined in two-dimensions (2D) on a square lattice by Villain\cite{Villain}, this model has been generalized in three dimensions (3D) as shown  in Fig. \ref{fig:SCFF} by Blankschtein et al.\cite{Blankschtein}. A detailed description of the model will be presented in section \ref{Model}. The nature of the phase transition in the classical XY\cite{Ngo2010} and Heisenberg\cite{Ngo2010a} spin models has been recently investigated.  It was shown that it is a first-order transition putting an end to a 25-year long controversial issue\cite{Diep85c,Diep85b}. In this paper, we extend our study to the case of Ising spin model.

In Section \ref{Model} we describe the model and give some technical details of the Wang-Landau
(WL) methods as applied in the present paper.  Section \ref{Res} shows our
results.  Concluding remarks are given in section \ref{Concl}.

\section{Model and Wang-Landau Method}\label{Model}

\begin{figure}
\begin{center}
\includegraphics{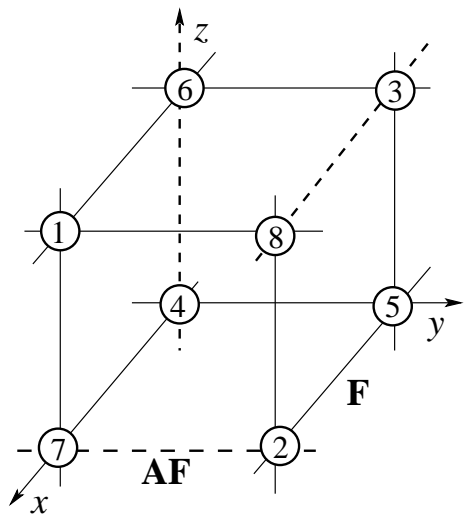}
\caption{Fully frustrated simple cubic lattice.  Discontinued (continued) lines denote antiferromagnetic (ferromagnetic) bonds.} \label{fig:SCFF}
\end{center}
\end{figure}

The model shown  in Fig. \ref{fig:SCFF} has been previously called "fully frustrated simple cubic lattice" (FFSCL) by Blankschtein et al.\cite{Blankschtein}.
The Hamiltonian is given by
\begin{equation}
{\cal H} = -\sum_{(i,j)} J_{ij}\mathbf{S}_i.\mathbf{S}_j,
\end{equation}
where $\mathbf{S}_i$ is the Ising spin of values $\pm 1$ at the lattice site $i$, $\sum_{(i,j)}$ is made
over the NN spin pairs $\mathbf{S}_i$ and $\mathbf{S}_j$ with  interaction $J_{ij}$.
We take $J_{ij}=-J$ ($J>0$) for antiferromagnetic bonds indicated  by discontinued lines in Fig. \ref{fig:SCFF}, and $J_{ij}=J$ for ferromagnetic bonds indicated by continued lines.  The 2D Villain's model has been intensively studied with Ising model\cite{Villain,deSeze} and XY spin model due to its application in arrays of planar Josephson's junctions\cite{Berge,Lee91,Diep98}.

\normalsize

 Let us recall some results on the present model. The GS degeneracy is infinite due to the fact that each face of the cube is frustrated, there is thus an infinite number to arrange the spins in an infinite crystal.  Note that ferromagnetic state is one of the GS spin configurations. In an early MC study\cite{Diep85a}, it has been shown that as the temperature $T$ increases, the system selects the long-range ferromagnetic state at low $T$ but goes to a partially disordered phase where two of the 8 sublattices of the cube are disordered.   The passage to this phase does not have the characteristics of a phase transition. The specific heat shows a "shoulder" at $T\simeq 0.5$ (in unit of $J/k_B$), far below the transition temperature for the whole system occurring at $T\simeq 1.345$.  Note that the nature of the low-$T$ ordering of the present model is still not elucidated. However, in 1987 we have shown\cite{Diep1987} in an exactly solved 2D model that a partial disorder can coexist with an order at equilibrium.  Therefore, we believe that the early observation of two disordered sublattices in an ordered phase may have the same origin rooted in the frustration and in the order selection by entropy\cite{Villain,deSeze,Villain80}.  The shoulder of the specific heat may turn out to be a true phase transition.  This point has to be checked with careful MC simulations using very large lattice sizes. This is a formidable task,  but it is not the purpose of this work.   In the present work, we concentrate our attention on the nature of the overall phase transition occurring at a higher temperature.  Using the Landau-Ginzburg-Wilson theory,  Blankschtein et al.\cite{Blankschtein}have found a weak first-order transition.  Our previous work in 1985 using a standard MC algorithm with short runs and small lattice sizes permitted by the computer capacity at that time\cite{Diep85a} show a second-order transition with an unusual critical properties in contradiction with the prediction of  Blankschtein et al.   In the light of new results on frustrated systems obtained  not only by new efficient MC algorithms but also  by today's huge computer capacity\cite{Ngo08,Diep2008,Ngo2010,Ngo2010a}, we study this problem again in order to get a definite answer to that question.


For weak first-order transitions, MC simulations with the standard Metropolis algorithm cannot give results with good precision even with the use of large sizes and long runs. This is because the algorithm does not allow us, among other difficulties, to easily sample rare microscopic states.    Wang and Landau\cite{WL1} have recently proposed a MC algorithm  which allowed to study classical statistical models with difficultly accessed microscopic states. In particular, it permits to detect  with efficiency weak first-order transitions\cite{Ngo08,Diep2008,Ngo2010} The algorithm uses a random walk in energy space in order to obtained an accurate estimate for the density of states $g(E)$ which is defined as the number of spin configurations for any given $E$. This method is based on the fact that a flat energy histogram $H(E)$ is produced if the probability for the transition to a state of energy $E$ is proportional to $g(E)^{-1}$.  At the beginning of the simulation, the density of states (DOS) is set equal to one for all possible energies, $g(E)=1$.
We begin a random walk in energy space $(E)$ by choosing a site randomly and flipping its spin with a probability
proportional to the inverse of the temporary density of states (DOS). In general, if $E$ and $E'$ are the energies before and after a spin is flipped, the transition probability from $E$ to $E'$ is
\begin{equation}
p(E\rightarrow E')=\min\left[g(E)/g(E'),1\right].
\label{eq:wlprob}
\end{equation}
Of course, to enhance the possibility to access to rare states, some tricks have been devised.
Each time an energy level $E$ is visited, the DOS is modified by a modification factor $f>0$ whether the spin flipped or not, i.e. $g(E)\rightarrow g(E)f$.
  At the beginning of the random walk, the modification factor $f$ can be as large as $e^1\simeq 2.7182818$. A histogram $H(E)$ records the number of times a state of energy $E$ is visited. Each time the energy histogram satisfies a certain "flatness" criterion, $f$ is reduced according to $f\rightarrow \sqrt{f}$ and $H(E)$ is reset to zero for all energies. The reduction process of the modification factor $f$ is repeated several times until a final value $f_{\mathrm{final}}$ which close enough to one. The histogram is considered as flat if
\begin{equation}
H(E)\ge x\%.\langle H(E)\rangle
\label{eq:wlflat}
\end{equation}
for all energies, where $x\%$ is chosen between $70\%$ and $95\%$
and $\langle H(E)\rangle$ is the average histogram.

The  WL method  has been applied  to our spin models with success in our recent papers\cite{Ngo08,Diep2008,Ngo2010}.  We  emphasize that for efficiency, we consider here a multi subinterval energy scale within an energy range of interest\cite{Schulz,Malakis}
$(E_{\min},E_{\max})$ which covers not all possible energies of the system but all energies in the region will will use in applications. We divide this energy range to $R$ subintervals, the minimum energy of the $i-th$ subinterval is $E^i_{\min}$ ($i=1,2,...,R$), and the maximum  is $E^i_{\max}=E^{i+1}_{\min}+2\Delta E$,
where $\Delta E$ can be chosen large enough for a smooth boundary between two subintervals. The WL
algorithm is used to calculate the relative DOS of each subinterval $(E^i_{\min},E^i_{\max})$ with a flatness criterion $x\%=95\%$.
Note that we reject a spin flip and do not update $g(E)$ and the energy histogram $H(E)$ of
the current energy level $E$ if the spin-flip trial would result in an energy outside the energy segment.
The DOS of the whole range is obtained by joining the DOS of each
subinterval $(E^i_{\min}+\Delta E,E^i_{\max}-\Delta E)$.

The thermodynamic quantities\cite{WL1,brown} can be evaluated by
\begin{eqnarray}
\langle E^n\rangle &=&\frac{1}{Z}\sum_E E^n g(E)\exp(-E/k_BT)\label{E}\\
C_v&=&\frac{\langle E^2\rangle-\langle E\rangle^2}{k_BT^2}\label{CV}\\
\langle M^n\rangle &=&\frac{1}{Z}\sum_E M^n g(E)\exp(-E/k_BT)\\
\chi&=&\frac{\langle M^2\rangle-\langle M\rangle^2}{k_BT}
\end{eqnarray}
where $Z$ is the partition function defined by
$Z =\sum_E g(E)\exp(-E/k_BT)$.
The canonical distribution at a temperature $T$ can be calculated simply by
$P(E,T) =\frac{1}{Z}g(E)\exp(-E/k_BT)$.

The simulations have been carried our on a rack of several hundreds of 64-bit CPU. For a given size $L$, the calculation takes, depending on $L$,  from a few weeks to several months to have the required histogram flatness.

\section{Results}\label{Res}

We have started the simulations from the system linear size $L=60$ (the system size is $L^3$).  But only from $L=90$ that a sign of first-order transition appears. Therefore, we use extremely large sizes up to 180. Periodic
boundary conditions are used in the three directions.  $J=1$ is
taken as the unit of energy in the following.

\begin{figure}
\begin{center}
\includegraphics{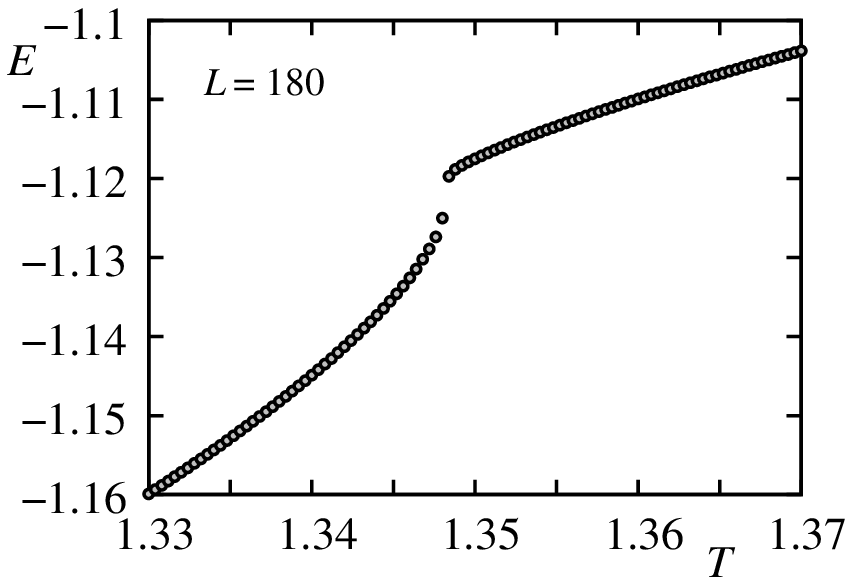}
\includegraphics{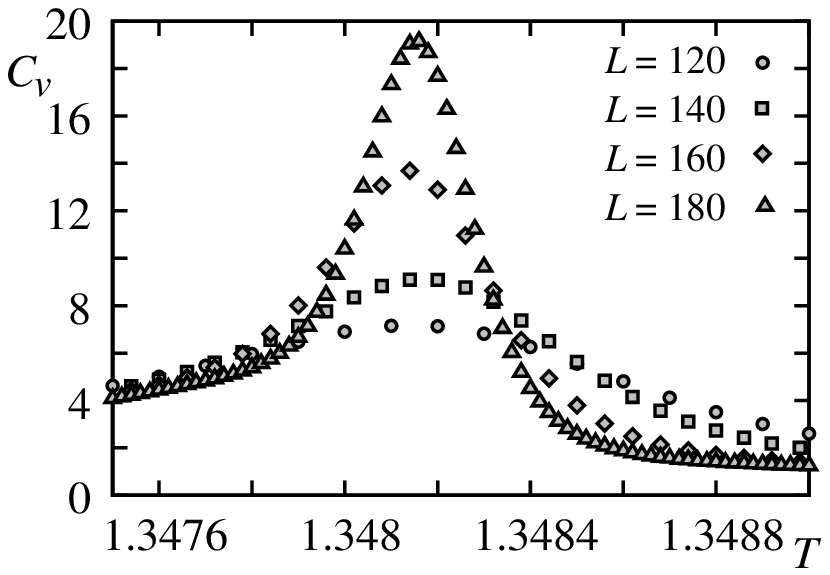}
\caption{Energy per spin $E$ versus temperature $T$ at the lattice size $180^3$ (upper figure) and specific heat per spin $C_V$
 versus $T$ for  lattice sizes $120^3$, $140^3$, $160^3$, $180^3$ (lower figure).  See text for comments.} \label{fig:EC}
\end{center}
\end{figure}

We show in Fig. \ref{fig:EC} the energy per spin and the specific heat, for $L=180$, using the flat histogram obtained with WL method.  Several remarks are in order:

i) the energy at the largest size shows a 'pseudo" discontinuity at the transition temperature $T_C\simeq 1.34814$.  We will see below that this discontinuity is confirmed by the double-peak energy histogram at this temperature,

ii) the specific heat shows a very strong size dependence.  It should be noted that the specific heat is calculated from the fluctuation of the  energy of the system at a given $T$ [see Eq. (\ref{CV})], not
by the derivative of $E$ with respect to $T$.  Therefore, when the energy has a discontinuity at $T_C$ with two levels $E_1$ and $E_2$, the average energy is $E=(E_1+E_2)/2$. It is the fluctuations of $E$ which gives rise to $C_V$, and we will not see a delta-like function should $C_V$ is calculated by the energy derivative.  This is the reason why in standard MC simulations with the Metropolis algorithm, we do not see discontinuity in energy for weak first-order transition (what is sorted out of the simulation is an average energy). So, an energy histogram is really needed if we want to see weak first order.

\begin{figure}
\begin{center}
\includegraphics{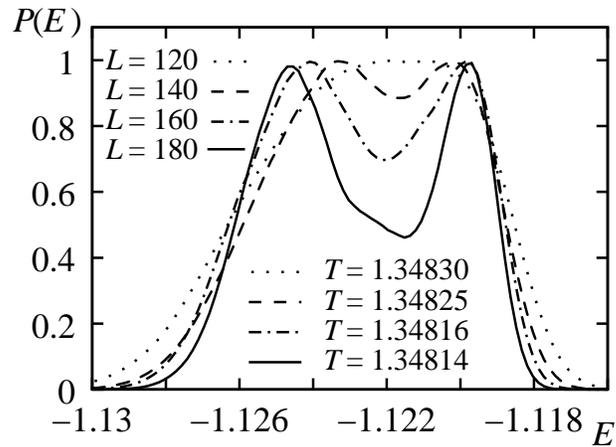}
\caption{Energy
histogram for several sizes $120^3$,  $140^3$, $160^3$, $180^3$ at $T_C$ indicated for each linear size on the
figure.}\label{fig:PE}
\end{center}
\end{figure}

The energy histogram can be realized directly in the old fashion manner by measuring the system energy at a given $T$\cite{Swendsen}.  However, when relaxation between rare states are very slow, we need the temperature-independent WL flat histogram technique as described above.    We show the WL result 
in Fig. \ref{fig:PE}. As seen, for $L=120$, the energy histogram begins to show a sign of the double-peak structure. The dip between the two maxima becomes deeper with increasing size.  Note that a "true" discontinuity happens only when the dip comes down to $E=0$.  This requires sizes much larger than $L=180$.  But for our present purpose, we need not to study sizes larger than $L=180$.

We note that the distance between the two peaks, i. e.
the latent heat, increases with increasing size and reaches $\simeq 0.005$ for $L=180$.  This is very small compared to the value $\simeq 0.03$ for  the XY case at $L=48$, and  to $\simeq 0.0085$ for the Heisenberg case at $L=90$.  The smallness of the latent heat in the present Ising case explains why one should go to an extremely large lattice size to detect the first-order transition.

\normalsize

\normalsize

\begin{figure}
\begin{center}
\includegraphics{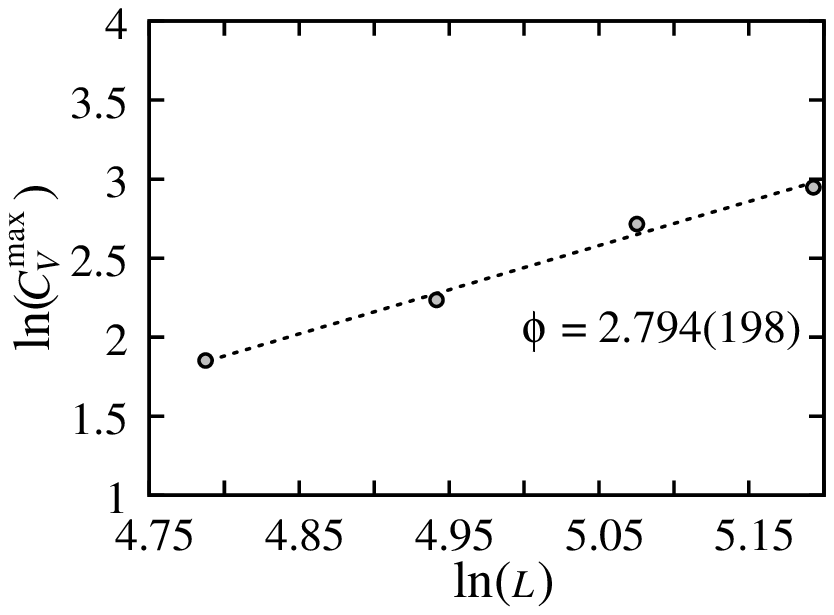}
\caption{Maximum of the specific heat $C_V^{max}$ versus $L$ in the $\ln-\ln$ scale.  The straight line is a mean least square fit.  The slope is $\phi=2.794(198)$.   Note that the specific heat shown in Fig. \ref{fig:EC} has been calculated from the fluctuations of the energy.}\label{fig:CVmax}
\end{center}
\end{figure}

Let us show in Fig. \ref{fig:CVmax}  the maximum of $C_V$  versus $L$ in a $\ln-\ln$ scale, we find a straight line within statistical errors (by a mean least-square fit) with a slope equal to $\phi= 2.794\pm 0.198$.  This means that $C_V^{max}=AL^\phi$ where $A$ is a constant and $\phi$ very close to the system dimension $d=3$.  The fact that $C_V^{max}$ is proportional to the system volume gives another strong signature of a first-order transition.

The weak first-order transition found here is thus in agreement with the Landau-Ginzburg-Wilson theory\cite{Blankschtein}. This is rather surprising because in other frustrated systems such as the STA mentioned in the Introduction, the renormalization group with low-order developments in $\epsilon$ did not work properly.

\normalsize

\section{Concluding Remarks}\label{Concl}

We have showed in this paper the results obtained by the  WL flat energy-histogram technique on the phase transition in the Ising fully frustrated simple cubic lattice.
We found that the transition is clearly of first order. Note that the first-order character is so weak that it has been observed only at extremely large lattice sizes. This finding shows that early studies using standard MC algorithm with short runs and much smaller sizes\cite{Diep85a} are not correct.  Our result confirms the prediction by the Landau-Ginzburg-Wilson analysis\cite{Blankschtein} putting an end to an uncertainty which has lasted for 25 years.
Together with our recent results\cite{Ngo2010,Ngo2010a}, we conclude that the fully frustrated simple cubic lattice undergoes a first-order transition for Ising, XY and Heisenberg spin models.
It is worth to mention that several other frustrated systems also show a first-order transition such as helimagnets\cite{Diep89a}, FCC\cite{Diep89} and HCP\cite{Diep92}  antiferromagnets.

This study shows that one has to be very careful in studying  complex systems by MC simulations: in some cases such as the one studied here,  sizes  as large as  $80^3$
are still not sufficient to get a correct conclusion.    Recent large-scale
MC simulations using special-purpose algorithms such as the WL technique have allowed us to settle several long-standing controversial questions\cite{Ngo08,Diep2008,Ngo2010,Ngo2010a}.

\section*{Acknowledgments}
One of us (VTN) would like to thank the University of Cergy-Pontoise for a financial support during the course of this work. He is grateful to Nafosted of Vietnam National Foundation
for Science and Technology Development, for support (Grant No. 103.02.57.09). He also thanks the NIMS (National Institute for Mathematical
Sciences, Korea) for hospitality and financial support.

\end{document}